\documentclass[twoside]{article}
\usepackage{fleqn,espcrc2,epsf}
\usepackage{graphicx}

\newcommand{\AmS}{{\protect\the\textfont2
  A\kern-.1667em\lower.5ex\hbox{M}\kern-.125emS}}

\title{Introducing Dynamical Triangulations to the Type IIB Superstrings}

\author{S. Oda\address{High Energy Accelerator Research Organization (KEK), 
        Tsukuba 305-0801, Japan}%
        \thanks{Presented by S.Oda}
        and 
        T. Yukawa\address{Coordination Center for Research and Education, 
	The Graduate University for Advanced Studies, 
	Miura-gun, Kanagawa 240-0193, Japan}$^{,a}$}
       
\begin{document}

\begin{abstract}
In order to consider non-perturbative effects of superstrings,
we try to apply dynamical triangulations to the type IIB superstrings.
The discretized action is constructed from the type IIB matrix model
proposed as a constructive definition of superstring theory.
The action has the local N=2 supersymmetry explicitly,
and has no extra fermionic degrees of freedom.
We evaluate the partition function for some simple configurations
and discuss constraints required from the finiteness of partition functions.
\vspace{1pc}
\end{abstract}

\maketitle

\section{Motivation}
From recent developments, the universal superstring picture arose,
namely the five types of superstrings and the M-theory are 
different representations of a single fundamental theory.
It is expected to be a constructive definition of superstrings.
The matrix model defined on the basis
of the 10-dimensional super Yang-Mills theory is considered to be
one of such constructive superstring theories
\cite{BFSS,IKKT,IT}.
There have been some attempts to introduce the dynamical
triangulations in superstrings as a lattice regularization
\cite{Siegel_random_GS,Ambhorn_random_simulation}.
In this study, we consider the type IIB superstring model
derived as the large $N$ limit of the $N \times N$ matrix theory
\cite{Random_IIB}.
The advantage of the type IIB superstring is that
the action has the local N=2 supersymmetry explicitly,
and has no extra fermionic degrees of freedom.

\section{Type IIB random superstrings}
The action of the type IIB matrix model is expressed 
in the semi-classical (i.e. large matrix size) limit by
\begin{eqnarray}
S \!\! = \!\! \int \!\! d^2 \sigma \sqrt{g} \left[
    \frac{1}{4} \{ X^{\mu}, X^{\nu} \}^2 \!\!
  - \frac{i}{2} \bar{\theta} \Gamma_{\mu} \{ X^{\mu}, \theta \} 
  \right] \!\!,     
\end{eqnarray}
where 
the Poisson brackets $\{ \;, \; \}$ are defined by
\begin{eqnarray}
\{X,Y\} \equiv \frac{1}{\sqrt{g}}\epsilon^{ab}\partial_a X \partial_b Y.
\end{eqnarray}
Here, $X^{\mu}$ are the $d$-dimensional space-time coordinates and
$\theta$ are anti-commuting spinor coordinates.
Also, $g$ is the absolute value of determinant of the world-sheet metric
($g_{ab}$), and $\epsilon^{ab}$ is the anti-symmetric tensor.
$S$ has the local $N=2$ supersymmetry for $d=3,4,6$ and $10$ cases,
and the $N=2$ super transformations are given by
\begin{eqnarray}
& & \left\{ 
\begin{array}{lcl}
\delta^{(1)}X^{\mu}   & = & i \bar{\epsilon}_1 \Gamma^{\mu} \theta \\
\delta^{(1)}\theta \: & = & - \frac{1}{2} \{X^{\mu},X^{\nu}\}
                                          \Gamma_{\mu \nu} \epsilon_1,
\end{array}
\right. \\
& & \left\{ 
\begin{array}{lcl}
\delta^{(2)}X^{\mu}   & = & 0      \\
\delta^{(2)}\theta \: & = &  \epsilon_2,
\end{array}
\right.                            \label{eq:Schild_con_SUSY_tr2}
\end{eqnarray}
where $\Gamma^{\mu \nu}$ is an antisymmetric tensor defined by
$\Gamma^{\mu \nu} = \frac{1}{2} [ \Gamma^{\mu}, \Gamma^{\nu} ]$.

In order to obtain the discretized action corresponding to $S$,
we perform triangulations of the world-sheet by equilateral triangles.
By summing over contributions from each triangle with three
vertices $i$, $j$ and $k$,
the discretized type IIB superstring action is given by
\begin{eqnarray}
S \sim
& & \hspace*{-6mm}
      \frac{\triangle}{2} \!\! \sum_{\langle ijk \rangle} \left[
              \frac{1}{4} \{ X^{\mu}, X^{\nu} \}_{\langle ijk \rangle}^2   
                                 \right. \label{eq:dis_action_B_Poisson}  
                                                \nonumber \\
 &  & 
\hspace*{-10mm}
      \left. 
        - \frac{i}{4} \bar{\theta}_{\langle ijk \rangle} \Gamma^{\mu} 
                             \{ X_{\mu}, \theta \}_{\langle ijk \rangle}
       \!\! - \!\! \frac{i}{4} \{ \bar{\theta} , X_{\mu} \}_{\langle ijk \rangle} 
                            \Gamma^{\mu} \theta_{\langle ijk \rangle}
                                                  \!  \right]
                                        \label{eq:dis_action_F_Poisson}  
                                                \nonumber \\
= \hspace*{3mm}
& & \hspace*{-9mm} \frac{1}{4 \triangle} \sum_{\langle ijk \rangle}  
        \left[ 
     - \frac{1}{4} \{ (X_{ij}^2)^2 + (X_{jk}^2)^2 + (X_{ki}^2)^2 \} 
                                            \right. \nonumber \\ 
& &  \hspace*{9mm}
     + \frac{1}{2} \{ X_{ij}^2 X_{jk}^2 
       + X_{jk}^2 X_{ki}^2 + X_{ki}^2 X_{ij}^2 \} 
                        \nonumber    \label{eq:Schild_Dbos_action} \\
& & \hspace*{3mm} 
     - \frac{\triangle}{3} \{ \Omega_{ij} \!\! \cdot \!\! 
                              (X_{jk} \!\! - \!\! X_{ki}) 
                \! + \! \Omega_{jk} \!\! \cdot \!\! (X_{ki} \!\! - \!\! X_{ij})
                                                        \nonumber \\
& & \hspace*{25mm} \left.
                \! +    \Omega_{ki} \!\! \cdot \!\! (X_{ij} \!\! - \!\! X_{jk}) 
                             \} \frac{}{}     \right].
                               \label{eq:Schild_Dfer_action} 
\end{eqnarray}
Here, 
$\theta_{\langle ijk \rangle} = \frac{1}{3} (\theta_i + \theta_j + \theta_k)$
is the average strength of the fermionic field for the triangle
$\langle ijk \rangle$,
and we denote 
$X_{ij}^{\mu} \equiv X_i^{\mu} - X_j^{\mu}$ and
$\Omega_{ij}^{\mu} \equiv \frac{i}{2} ( \bar{\theta}_i \Gamma^{\mu} \theta_j 
                      - \bar{\theta}_j \Gamma^{\mu} \theta_i )$.
$X^{\mu}_i$ and $\theta_i$ are bosonic and fermionic fields 
on a vertex $i$, respectively,
and $\triangle$ is twice the area of elementary triangle.
The discretized Poisson brackets are defined by
\begin{eqnarray}
\{A,B\}_{\langle ijk \rangle} 
& \equiv & \frac{1}{\triangle} f_{ijk} (A_{jk}B_{ki} - B_{jk}A_{ki}),
                                             \label{eq:Schild_dis_Poisson}
\end{eqnarray}
where
\begin{eqnarray}
f_{ijk}\hspace{-4mm} & = & \hspace*{-4mm}
\left\{
\begin{array}{cl}
+1 & (i,\; j, \;k) \; {\rm counterclockwise} \\
-1 & (i,\; j, \;k) \; {\rm clockwise}.             
\end{array}
\right. 
\end{eqnarray}
%
$S$ is invariant 
under the following descretized super transformations;
\begin{eqnarray}
& & \left\{
\begin{array}{lcl}
\delta^{(1)}X_i^{\mu} & = & i \bar{\epsilon}_1 \Gamma^{\mu} \theta_i \\
\delta^{(1)}\theta_{\langle ijk \rangle} \: 
& = & - \frac{1}{2} \{ X^{\mu},X^{\nu} \}_{\langle ijk \rangle} 
                                   \Gamma_{\mu \nu} \epsilon_1, 
                                    \label{eq:Schild_dis_SUSY_tr1} 
\end{array} 
\right. \\
& & \left\{
\begin{array}{lcl} 
\delta^{(2)}X_i^{\mu} & = & 0   \\ 
\delta^{(2)}\theta_i \: & = & \epsilon_2.   \label{eq:Schild_dis_SUSY_tr2}
\end{array}
\right. 
\end{eqnarray} 
Since the discretized Poisson bracket is constant on each triangle,
the variation of the fermionic field 
$\delta^{(1)}\theta_{\langle ijk \rangle}$ is defined on the triangle.
Then we must examine the one-to-one correspondence between
the variations of fermionic variables on
triangles and on vertices for the local supersymmetric invariance.
In order to check the correspondence,
we count the number of independent relations
among the $N_2 = 2 (N_0 - \chi)$ equations,
\begin{eqnarray}
\delta^{(1)} \theta_{\langle ijk \rangle} & = & \frac{1}{3} \left(
\delta^{(1)} \theta_i + \delta^{(1)} \theta_j + \delta^{(1)} \theta_k \right),
                                 \label{Eq:one-to-one}
\end{eqnarray}
with the Euler characteristic $\chi$.
According to the attempt for numerical solutions of eq.~(\ref{Eq:one-to-one}),
we find that there dose not exist the one-to-one correspondence only 
for those configurations which have extra spatial symmetry.
Since these configurations are expected to possess degeneracy,
we should include extra measure factors to impose the
one-to-one correspondence.

\section{Partition functions and numerical results}
Quantization is carried out using a standard path-integral method.
The path integration over $\sqrt{g}$ is interpreted 
as the sum over dynamical triangulations.
For the type IIB superstring,
the discretized partition function with a fixed number of vertices is given by
\begin{eqnarray}
Z(N_0) \! & = & \! \sum_{{\cal T}} \!\!
       \! \frac{1}{\cal S(T)} \!\! \int \!\! \prod_{i=0}^{N_0-1} \!\!\!
          d X_i d \theta_i d \bar{\theta}_i e^{- S_B - S_F} \nonumber \\
& & \hspace*{20mm} \times \; 
          \delta (X_0) \delta (\bar{\theta}_0) \delta (\theta_0),
\end{eqnarray}
where ${\cal S(T)}$ is a symmetry factor of triangulations (${\cal T}$) 
to take into account the degeneracy.
It is noted that the modes associated with 
the translational invariance of the action for bosonic fields and 
the $\delta^{(2)}$ supersymmetry for fermionic fields is eliminated by
imposing $X_0=0$ and $\theta_0 = \bar{\theta}_0 = 0$.
However, there exist special configurations which give 
the zero fermion determinant.
They don't have the one-to-one correspondence 
between triangles and vertices resulting fermionic zero modes.
In order to obtain the non-zero fermion determinant,
we include extra constraint to kill dependent fermionic fields as
$\prod_k^{{\tiny N_0^{dep.}}} \delta(\bar{\theta}_k) \delta(\theta_k)$,
where $N_0^{dep.}$ is the number of dependent relations.

We will evaluate the partition function of 
a tetrahedron ($N_0=4$) configuration as shown by solid lines in Fig.~1.
Their integration measures are defined by 
\begin{eqnarray}
\begin{array}{ccl}
d X_1 & = & x^{d-1} dx d \Omega_{d-1},   \\
d X_2 & = & dy z^{d-2} dz d \Omega_{d-2},  \\
d X_3 & = & \prod_{i=1}^d dx_i.          
\end{array} 
\end{eqnarray}
By integrating out the fermionic coordinates,
%
the partition function is given by
\begin{eqnarray}
Z(4) \!
\! & \propto & \!\! \int_0^{\infty} \!\! x^{d-1} z^{d-2} dx dz 
\!\! \int_{-\infty}^{\infty} \!\! dy
      \prod_{i=1}^{d} dx_i e^{- S_B} \nonumber \\
& & \times \left[ (x_3^2 + \cdots + x_d^2) x^2 z^2 
                       \right]^{\frac{k}{2}} \nonumber \\
& \propto & \int_0^{\infty} dX X^{\frac{1}{2}(d+k-2)}
            e^{- \frac{4}{3} X^2} \nonumber \\
& & \hspace*{-6mm} \times \!\! \int_0^{\frac{\pi}{2}} \!\!\!\! d \theta 
            (\cos \theta)^{\frac{1}{2}(d+k-4)}
            (\sin \theta)^{\frac{1}{2}(d+k-6)} \!\! ,
                               \label{eq:Schild_tet_Z_last}
\end{eqnarray}
where
\begin{eqnarray}
S_B \hspace*{-2mm} & \cong & \hspace*{-2mm}
 x^2 z^2 + (z x_1 - y x_2)^2 + x_2^2 x^2  \nonumber \\
& & \hspace*{-4mm} + \{ (y-x) x_2 - (x_1 - x) z \}^2 \nonumber \\
& & \hspace*{-4mm} + 2\{ x^2 + y^2 + z^2 -xy \}
        (x_3^2 + \cdots + x_d^2).
\end{eqnarray}
We have defined $X^2 \equiv x^2 z^2$, 
$\tan \theta \equiv \frac{2}{\sqrt{3}} \frac{z}{x}$,
and $\Gamma^{\mu}$ are $k \times k$ matrices. 
The condition in which the partition function is finite is
$d+k > 4$.
In the case of a tetrahedron,
even the partition function of the bosonic string becomes finite, 
when $d > 4$.
However, the situation is not so simple for larger size configurations.
We have found numerically that the partition function
of the bosonic string exhibits a spike singularity for large $N_0$
in the case of $d=5$.

In the case of a configuration with an additional vertex as shown in Fig.~1,
where links of three triangles added by the 5th vertex is shown in dotted lines,
we can obtain two conditions as
$k > 0$ and $2d+k > 6$.
The condition $k > 0$ implies that
contributions from fermionic fields are necessary
for the finiteness of the partition function.

\begin{figure}[htb]
\epsfxsize=7cm \epsfysize=4cm
\centerline{\epsfbox{5-vertex.eps}}
\caption{%
Solid lines denote a tetrahedron configuration,
and adding dashed lines they denote a configuration with 5 vertices.
}
\label{Fig:5_vertex}
\epsfxsize=6cm \epsfysize=5cm
\centerline{\epsfbox{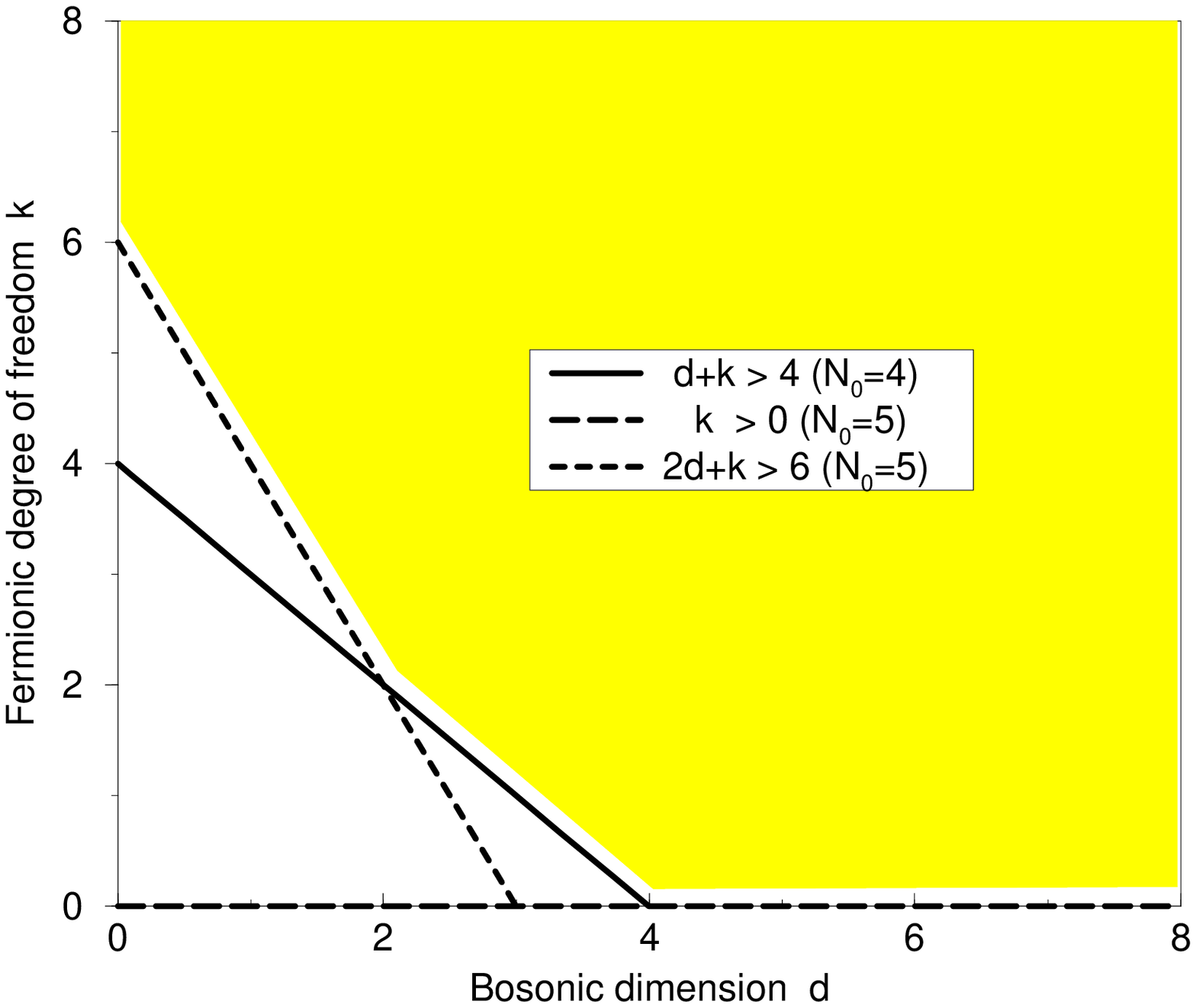}}
\vspace*{-8mm}
\caption{%
Constraint conditions from the analytic calculations for
the partition functions of $N_0 = 4$ and 5 configurations.
In the shadowed region, these three constraints are satisfied.
}
\label{Fig:Constraint}
\end{figure}

\section{Summary}
We have proposed the discretized type IIB superstring action,
and have shown that it is invariant under the local N=2 super transformations.
In the numerical study, the partition function of the bosonic strings
exhibits a spike singularity for large configuration,
and spike configurations dominate.
From the study with $N_0=5$ configuration,
we find that it is essential for the finiteness of the partition function
to introduce fermionic fields.
For further study, we need to find a powerful method to calculate the 
fermion determinants in a larger system.

\section*{Acknowledgements}
We wish to give special thanks to Prof. H. Kawai, N. Ishibashi,
F. Sugino and M. Sakaguchi for many helpful discussions.


\end{document}